\documentclass[12pt,epsf]{article} 
\usepackage{tabularx}
\usepackage{amssymb,amsmath,amscd}
\usepackage{amsfonts}
\usepackage{epsfig,float,wrapfig}
\newcommand{\nc}{\newcommand}
\def\frac#1#2{{\textstyle {#1 \over #2}}}

\nc{\beq}{\begin{equation}}
\nc{\eeq}{\end{equation}}
\nc{\beqa}{\begin{eqnarray}}
\nc{\eeqa}{\end{eqnarray}}
\nc{\lsim}{\begin{array}{c}\,\sim\vspace{-21pt}\\< \end{array}}
\nc{\gsim}{\begin{array}{c}\sim\vspace{-21pt}\\> \end{array}}
\nc{\appA}{}
\nc{\appB}{}
\nc{\appC}{}
\nc{\appD}{}
\nc{\appE}{}
\nc{\Eqr}[1]{(\ref{#1})}
\nc{\mysection}[1]{\setcounter{equation}{0}\section{#1}}

\nc{\myappendix}[1]{\section*{#1}\setcounter{equation}{0}}

\def\bra{\langle }
\def\ket{\rangle }

\def\dk{d\bar{k}}
\def\dr{d\bar{r}}
\def\ds{d\bar{s}}
\def\ddp{d\bar{p}}
\def\dq{d\bar{q}}

\def\&{and}

\def\D{ {\cal D} }
\def\det{\operatorname{det}}
\def\Tr{\operatorname{Tr}}
\def \Eslash {E \kern-.5em\slash}

\def\com#1#2#3{          {\it Comm. Math. Phys. }{\bf #1}, #2 (#3)}

\def\imp#1#2#3{           {\it Int. J. Mod. Phys. }{\bf #1}, #2 (#3)}
\def\mpl#1#2#3{           {\it Mod. Phys. Lett. }{\bf #1}, #2 (#3)}

\def\nc#1#2#3{           {\it Nuovo Cim.  }{\bf #1}, #2 (#3)}
       
\def\np#1#2#3{           {\it Nucl. Phys. }{\bf #1}, #2 (#3)}
\def\pl#1#2#3{           {\it Phys. Lett. }{\bf #1}, #2 (#3)}

\def\pr#1#2#3{           {\it Phys. Rev. }{\bf #1}, #2 (#3)}
\def\prep#1#2#3{         {\it Phys. Rep. }{\bf #1}, #2 (#3)}

\begin{document}
\begin{titlepage}
\renewcommand{\thefootnote}{\fnsymbol{footnote}}
\begin{center}
\hfill
\vskip 1 cm
{\large \bf Path integral regularization of \\
 pure Yang-Mills theory  \\
}
\vskip 1 cm
{
  {\bf J. L. Jacquot}\footnote{jean-luc.jacquot@ires.in2p3.fr}
   \vskip 0.3 cm
   {\it IPHC-DRS, IN2P3-CNRS/Universit\'e de Strasbourg,\\
        23 rue du Loess,\\
        BP 28 - 67037 Strasbourg Cedex 2, FRANCE}\\ }
  \vskip 0.3 cm
\end{center}
\vskip .5 in
\begin{abstract}
In enlarging the field content of pure Yang-Mills theory to a cutoff dependent matrix valued complex scalar field,  we construct a vectorial operator, which is by definition  invariant with respect to   the gauge transformation of the Yang-Mills  field  and with respect to a Stueckelberg type gauge transformation of the  scalar field.
This invariant operator  converges to the original Yang-Mills field as the cutoff goes to infinity.
With the help of cutoff functions,  we construct with this invariant a regularized action for the pure Yang-Mills theory.
In order to be able to define  both  the gauge and scalar fields kinetic terms, other invariant terms are added  to the action.
Since the  scalar fields flat measure is invariant under the Stueckelberg type gauge transformation, we obtain a regularized gauge-invariant path integral for pure Yang-Mills theory that is   mathematically well defined.
Moreover, the regularized Ward-Takahashi identities describing the dynamics  of the gauge fields are exactly the same as the formal   Ward-Takahashi identities of the unregularized theory.\\[2mm]
PACS numbers: 11.15.Tk, 11.10.Lm, 11.10.Hi, 11.15.Ha
\end{abstract}

\end{titlepage}
\renewcommand{\thepage}{\arabic{page}}
\setcounter{page}{1}  
\setcounter{footnote}{0}
\mysection{Introduction}
To understand the dynamic of pure Yang-Mills (YM) theory \cite{FADDEEV}, in all the range of the energy scale, one needs  first a gauge-invariant regularization in four-dimensional spacetime.
In the seventies, Wilson has built a gauge-invariant regularization of YM theory by approximating the continuous four-dimensional spacetime as a discrete lattice  \cite{ROTHE}.
In this approach gauge invariance is obvious, but the continuous symmetries of the physical spacetime, such as Lorentz invariance are  clearly lost.
The continuous YM action is recovered when the lattice spacing goes to zero.
In this regime the theory is perturbative, and for each physical quantity, the result calculated in lattice gauge theory must match the result given in a perturbative continuous regularization, such as dimensional regularization \cite{COLLINS}.
For instance, as  the lattice spacing goes to zero, some integrals  which arise from   the diagrammatic expansion of the action of YM  lattice perturbation theory, like the  tadpole integrals,  become not regularized in the infrared (IR) and need an intermediate regularization  \cite{BECHER,CAPITANI}.
Therefore, since the matching between the discrete and continuous regularization is really not obvious, for the study of the evolution of  physical quantities with the energy scale, ranging from the IR to the ultraviolet (UV) domains, a continuous nonperturbative regularization in four-dimensions which will preserve both the spacetime symmetries and the gauge invariance of the theory, will be very appealing.

Apart from dimensional regularization, which is in fact purely perturbative, two different classes of  continuous  regularization were essentially developed as a starting point for a nonperturbative approach, since the advent of lattice regularization. 
1)In the method of higher covariant derivative supplemented by the introduction of  Pauli-Villars (PV)  regulator fields  \cite{FADDEEV,SLANOV}, which are necessary to regularize the one loop diagrams, it is the action which is regularized.
Because of Gribov  \cite{GRIBOV} ambiguities, in  this scheme  the regularized path integral has nevertheless only a perturbative meaning.
2)In the method of stochastic regularization  \cite{HALPERN1}, by construction it is not the YM action which is regularized with the help of covariant derivative, but the second order Schwinger-Dyson (DS) equations, which are  obtained after integrating out the Gaussian noise field.
Since in this scheme the quantization occurs without Faddeev-Popov ghosts, the regularized DS equations can in principle be used to study the nonperturbative aspects of YM theory.
However  in this scheme the renormalization program is not straightforward.
In fact, due to the lack of a regularized action, the regularized Ward-Takahashi (WT) identities can not be deduced from the onset, and  one must assume  \cite{HALPERN2} that they are identical to the unregularized ones.

In order to be able to apply exact renormalization group  techniques \cite{BAGN}, we require that the regularization scheme does not work only on Feynman diagrams, or DS equations, but regularize the YM path integral as a whole, explicitly and not only formally\footnote{For instance in the case of dimensional regularization  the path integral has only a perturbative meaning.}.
One method to regularize the path integral of a given theory, is to add a cutoff function in the quadratic term of the action, in order that the propagator becomes a rapidly decreasing function of the square of the momenta in momentum space.
This idea is one of the cornerstones of the construction of regularized path integrals which are suited for the derivation   of exact renormalization group  techniques  \cite{POLCHINSKI}.
In this approach, whereas the path integral that quantizises the theory is indeed well defined, the regularization breaks explicitly the internal local symmetries of the original action.
This is because the gauge transformation mixes different scales of the gauge field momenta.
Only recently  was it shown that this drawback can be circumvented by enlarging the fields content of the theory : The higher covariant derivative regularization is  applied to a spontaneously broken supergroup which embeds the original  physical non-Abelian gauge group  \cite{MORRIS0}.
The original YM theory is then  regularized by   higher covariant derivatives  supplemented by a finite set of PV regulator fields, if some preregularization is used  \cite{MORRIS1}.
But, because the   PV regulator fields introduced in the action have  negative norm by definition \cite{ZINN}, the path integral of the theory is  only defined formally and perturbatively. 

This paper shows that the path integral of $SU(3)$ YM theory can be regularized as a whole, in a  continuous and gauge-invariant way, by enlarging the field content of the theory by a finite set of complex scalar fields, smeared with cutoff functions.
In order to do so, we will follow  the  lines of reasoning which have allowed us to regularize the path integral of QED  \cite{JLJ} in a gauge-invariant manner. 
The action is constructed in its essence with  a  gauge-invariant building block  $\mathcal{A}_{\mu}^{Inv}$.
This vectorial operator of dimension one is a  function of the gauge field and of the smeared complex  scalar fields, and converges to the standard gauge field $A^{\mu}$ when the cutoff goes to infinity.
The variation of this vectorial operator under a local gauge transformation of the gauge field is compensated by a group transformation of the smeared complex scalar field which depends on the cutoff scale and which plays a role similar to that of Stueckelberg compensating fields \cite{RUEG}.
The regularization is implemented in the action by smearing  these invariants by smooth cutoff functions.
Then  all interaction vertices of  the regularized action are smeared with a cutoff function and the quadratic terms in the fields can be inverted to give rise to  gauge and scalar fields propagators which  are rapidly decreasing functions of the square of the momenta.
Since  the measure and the regularized action is invariant under both the gauge transformations of the vectorial and scalar fields, when  the integration over the scalar fields is being assumed implicitly, the regularized WT identities relative to the gauge field are exactly the same as those which can be deduced naively from the unregularized action without gauge fixing.

The paper is organized as follows.
Section II is devoted to the explicit construction of the regularized gauge-invariant action of the YM theory with the help of the invariant vectorial operator  $\mathcal{A}_{\mu}^{Inv}$.
The flat measure relative to the scalar fields is showed to be invariant under the gauge Stueckelberg type  transformation.
In Section III, in order to be able to invert the gauge vector kinetic term without breaking explicitly the whole gauge invariance of the regularized action, we add to the   action a gauge-invariant part  which is built with the invariant  operator $\mathcal{A}_{\mu}^{Inv}$ and  which is reminiscent of the standard covariant gauge fixing term.
Here the addition of a mass term for the real and imaginary parts of the complex scalar fields is required, so that the full action remains regularized.
In Section IV, we show  that  the mass term relative to the real part of the complex scalar field, that is not invariant under the   Stueckelberg  type of gauge transformation, is in fact harmless.
This implies that, after integration on the scalar fields,  the regularized WT identities  describing  the dynamics  of the YM gauge fields are exactly the same as the formal WT identities of the unregularized theory without gauge fixing.
\mysection{The regularized gauge-invariant action }
If  $A_{\mu}=A^{\mu}_{\alpha}T^{\alpha}$, is the $SU(3)$ color gauge vector boson, we define the matrix valued gauge-invariant operator  
\beq
\label{Ainv1}
\mathcal{A}_{\mu}^{Inv}=U^+A_{\mu}U -\frac{1}{2}(U^+\partial_{\mu}U -\partial_{\mu}U^+U),
\eeq
with the generators $T^{\alpha}$ related to  the Gell-Mann matrices by
\beq
\label{generator1}
T^{\alpha}=i\frac{\lambda_{\alpha}}{2}.
\eeq
Here, $U$ is a $3 \times 3$ dimensionless complex matrix  acting on the generators $T^{\alpha}$, and  hereafter an expression like  $\partial_{\mu}AB$ means that the partial derivative acts only on the factor $A$. 
By construction the vector $\mathcal{A}_{\mu}^{Inv}$ is anti-Hermitic and invariant under  the local transformation of $A^{\mu}$ and $U$ by  an arbitrary $SU(3)$ matrix
\beqa
\label {gaugetrans0}
 A_{\mu}~&\rightarrow &~ VA_{\mu}V^+ +\partial_{\mu}VV^+  \\
\label {gaugetrans1}
 U~ &\rightarrow & ~ VU.
\eeqa
The invariance of ${A}_{\mu}^{Inv}$ relies only on the property of unitarity of the matrix $V$.
Thus, the  unitarity \footnote{If $U$ were unitary, the definition  (\ref{Ainv1}) is that  of the gauge-invariant operator given in  \cite{DRAGON}.} of the matrix $U$  is in general not required.
In order to construct  a regularized gauge-invariant action, we start with the gauge-invariant expression
\beq
\label{Ainv2}
 F_{\mu \nu}^{Inv}(z)=\int dx~\left \{ \rho^{(-1)}_1 (z,x)\left (\partial_{\mu}{A}_{\nu}^{Inv} -\partial_{\nu}{A}_{\mu}^{Inv}\right ) -\rho_2 (z,x) [{A}_{\mu}^{Inv},{A}_{\nu}^{Inv}]\right \}.
\eeq
Here and hereafter we work in Minkowski space.
We choose the signature of the metric to be $ (1,-1,-1,-1) $ and the notation $ dx \equiv  d^4x $ and $ \dk \equiv  d^4k/(2\pi )^4 $.
In the  expression (\ref{Ainv2}) the ultraviolet  UV or infrared IR regularization are implemented through the real scalar functions
\beqa
\label{cutf1}
\rho_i (x,y)& =& \int \dk ~e^{-ik(x-y)}\rho_i (\frac{k^2}{\Lambda^2},\kappa ) \\
\label{cutf0}
\rho^{(-n)}_i (x,y)& =& \int \dk ~\frac{e^{-ik(x-y)}} {\rho_i^n (\frac{k^2}{\Lambda^2},\kappa )~ +~ \eta  \rho_i^{-n} (\frac{k^2}{\Lambda^2},\kappa )},
\eeqa
$\kappa $ and $\Lambda$ being respectively the IR and UV cutoff scales.
The term proportional to the function $\rho^{(-1)}_1$ will contribute to the kinetic part of the action, and hence   to the inverse of the propagator.
As a result,  when the path integral of the theory  is  expressed in momentum space, the infinitesimal real parameter $\eta$, which was  only introduced to give a  mathematical meaning to the Fourier transform  (\ref{cutf0}),  can be set to  zero .
In Euclidean momentum space, the cutoff functions $\rho_i (k )$ are rapidly decreasing functions of $k^2$  in both the IR and UV domains and verify the condition \footnote{Notice that there are a large number of cutoff functions which are valid in the  IR and UV domains and which converge to the  Dirac's $\delta $ function as the cutoff $\Lambda$ goes to infinity, whatever the value of $\kappa$. For instance, this is the case for the cutoff function $\frac{k^2}{k^2 + \frac{\kappa^4}{\Lambda^2}} e^{\frac{k^2}{\Lambda^2}}$.}
\beq
\label{limcutf1} 
\lim_{\Lambda \to \infty,\kappa  \to 0} \rho_i (k,\Lambda,\kappa  )~=~1.
\eeq
By  definition, the  functions (\ref{cutf1}) are  regularized forms of Dirac's $\delta $ function, i.e.
\beqa
\label{cutf3}
\lim_{\Lambda \to \infty,\kappa  \to 0}\rho_i (x,y)&=&\delta (x-y).  
\eeqa

The expression  (\ref{Ainv2}) is apparently quartic in the matrix $U$.
In order to simplify and to express the operator (\ref{Ainv2}) in terms of the usual YM stress tensor, we suppose  as a first step, that the  matrix $U$ is unitary.
Then  using the relation 
\beq
\label{diffU1}
\partial_{\mu}U^+U =-U^+\partial_{\mu}U
\eeq
(\ref{Ainv2}) becomes 
\beqa
\label{Ainv3}
 F_{\mu \nu}^{Inv}(z)&=&\int dx~\bigg \{ U^+ F_{\mu \nu}^{Reg}(z,x) U +(  \rho^{(-1)}_1  (z,x  ) -\rho_2  (z,x)  )\Big [ ( \partial_{\mu}U^+{A}_{\nu} U  +  U^+{A}_{\nu}\partial_{\mu} U  \nonumber \\
&&   - \partial_{\mu}U^+\partial_{\nu}U ) -  ( \mu \leftrightarrow \nu ) \Big ] \bigg \},
\eeqa
where  the regularized YM stress tensor is defined by the expression
\beq
\label{Ainv4}
  F_{\mu \nu}^{Reg}(z,x)= \rho^{(-1)}_1 (z,x)\left (\partial_{\mu}{A}_{\nu} -\partial_{\nu}{A}_{\mu}\right ) -\rho_2 (z,x) [{A}_{\mu},{A}_{\nu}].
\eeq
If one relaxes the condition of unitarity of the matrix $U$, one can check that the new expression (\ref{Ainv3}) of the operator  (\ref{Ainv2}) remains invariant under both transformations  (\ref{gaugetrans0}) and (\ref{gaugetrans1}).
Notice that in this case the relation (\ref{diffU1}) does not hold true, and hence the relations (\ref{Ainv2}) and (\ref{Ainv3}) define two distinct invariants.
We select the last expression (\ref{Ainv3}) as the building block of the regularized YM action.
In  the limit $\Lambda \to + \infty$ , $\kappa \to 0$, one can set the parameter $\eta$ to zero, and the right-hand side of  (\ref{Ainv3})  converges to the operator $U^+ F_{\mu \nu} U$.
The trace of the operator $ F_{\mu \nu}^{Inv}  F^{Inv\mu \nu} $ does converge to the standard, unregularized Lagrangian density of QCD if the matrix $U$ is unitary.

The next step consists in  choosing the form of the dimensionless  matrix $U$.
Since the matrix $U$  is needed to restore the gauge invariance in (\ref{Ainv3}), it can be expressed as  a functional of some scalar fields which will play the role of Stueckelberg compensating fields  \cite{RUEG,DRAGON}.

First to avoid any nonpolynomial interaction between the  Stueckelberg  fields, which will enter  the action we are looking for, second to have   the path integral measure  invariant under the transformation of  the  Stueckelberg  fields that induces the transformation  (\ref{gaugetrans1}), and third, so that the invariant operator  $\mathcal{A}_{\mu}^{Inv}$ converges to the standard gauge vector  $A_{\mu}$ as the cutoff $\Lambda$ goes to infinity, we  define the $3 \times 3$ dimensionless matrix $U$ as the linear combination
\beq
\label{Umatrix1}
U=I + \frac{1}{\Lambda} \phi.
\eeq
Here, the matrix $\phi$ is expressed in terms of the complex scalar fields $ \phi_a$ as
\beq
\label{Umatrix2}
\phi =  \phi_a T^a,
\eeq
and   by convention, we use Latin letters when the index runs from $0$ to $8$. 
In the definition  (\ref{Umatrix2}),  except the matrix $T^0$  which is given in term of the identity matrix $I$ as
\beq
\label{generator2}
T^0=\frac{i}{\sqrt{6}} I,
\eeq
 the eight remaining matrices  $T^{\alpha}$  are defined in (\ref{generator1}).
With the definition (\ref{generator2}), the matrices $T^a$  are  normalized as 
\beq
\label{generator3}
\Tr( T^a T^b)= -\frac{1}{2}\delta_{ab},
\eeq 
and the properties of the matrices (\ref{generator1}) are recalled in  Appendix A.
We do not impose any constraint on the nine complex fields $\phi_i$, so that the matrix $U$ is not unitary  in general\footnote{For a  unitary  matrix $U$, the invariant measure of the fields $\phi_i$ is  hard to define, as it is the case in the nonlinear sigma model \cite{ZINN}.}.
Due to its definition, this matrix is an element of the Lie algebra of the $U(3)$ unitary group.
Since the matrix $U$ transforms like (\ref{gaugetrans1}), the transformed matrix $\phi'$  verifies the equation
\beq
\label{gaugetrans2}
 \phi'= V \phi +\Lambda  (V -I  ).
\eeq
It is clear that the set of transformations (\ref{gaugetrans2}), acting on the matrix  $ \phi$  (\ref{Umatrix2}),  is a nonlinear representation of the $SU(3)$ group.
In fact, if $g$ is the group element of $SU(3)$ represented by the matrix $V$, the transformed field $ \phi'$ can be written as
\beq
\label{grouptrans1}
 \phi'~=~ ^g\phi ,
\eeq
and we have 
\beq
\label{grouptrans2}
 ^{g'} (  ^g\phi  )~ =~ ^{g'g}\phi ,
\eeq
with the identity group element $e$ represented by the identity matrix $I$, and the inverse group element $g^{-1}$  represented by the  matrix $V^{-1}$.

How does the measure of the scalar field  transform under the transformation  (\ref{gaugetrans2}) ?
As show in  Appendix B, if  $\Phi$ is the  nine-dimensional complex vector of components  $\phi_i$, the transformed fields $\phi_i'$ are given according to $\Phi$ by the vectorial relation
\beq
\label{Umatrix3}
\Phi' = S^{-1} T\left (V \right ) S \Phi + \Lambda S^{-1} T\left(V -I \right) C_I.
\eeq
Here  the  nine-dimensional matrix representation  $ T(V)$ of the  $SU(3)$ gauge group  and the constant    vector $C_I$ are respectively defined in  (\ref{Tmatrix0}) and  (\ref{Tmatrix1}), and  $S$   is  the   constant  matrix  defined by the relation   (\ref{Tmatrix5}). 

Then the  Jacobian of the transformation  (\ref{Umatrix3}) is given by
\beq
\label{Jacobmatrix0}
\left \vert \frac{\delta \phi_i' (x)}{\delta \phi_j (x')} \right \vert =\delta (x-x') \det \big ( S^{-1} T\left (V \right ) S \big )= \delta (x-x') ,
\eeq
because the determinant of $T\left (V \right ) $ is unity  (\ref{Tmatrix3}).
In the same manner, one can see that the Jacobian of the transformation of the Hermitic conjugate scalar fields $\phi_i^+$,  induced by the transformation  (\ref{gaugetrans2}), is also unity.
As a result, since the scalar fields  $\phi_i$ and  $\phi_i^+$ are independent variables the  measure
\beq
\label{scalarmeasure1}
\D \phi \D \phi^+ \equiv \prod_{i=0}^{i=8}\D \phi_i \prod_{i=0}^{i=8}\D \phi_i^+ ,
\eeq
is invariant under the transformation  (\ref{gaugetrans2}).  

Since the form of the matrix $U$ is now given by the expression (\ref{Umatrix1}), the invariant tensor  (\ref{Ainv3}) is recast in terms of the regularized YM stress tensor (\ref{Ainv4}) in the expression
\beq
\label{Ainv5}
 F_{\mu \nu}^{Inv}(z)=\int dx~\left \{  F_{\mu \nu}^{Reg}(z,x) +  \rho_2 \left (z,x \right ) G_{\mu \nu} + \left(  \rho^{(-1)}_1 \left (z,x \right ) -\rho_2 \left (z,x\right) \right ) H_{\mu \nu} \right \} ,
\eeq
where the antisymmetric tensors $G_{\mu \nu}$ and $H_{\mu \nu}$ are given in terms of the gluons fields $A_{\mu}$, the standard YM strength tensor $ F_{\mu \nu}$ and the complex scalar fields $\phi$ by
\beqa
\label{G1}
G_{\mu \nu}&=&\frac{1}{\Lambda} (  \phi^+ F_{\mu \nu} + F_{\mu \nu} \phi )  + \frac{1}{\Lambda^2} \phi^+  F_{\mu \nu} \phi \\
\label{H1}
H_{\mu \nu} & =& \big [ \frac{1}{\Lambda} \partial_{\mu}  ( \phi^+ A_{\nu} +  A_{\nu} \phi  ) +  \frac{1}{\Lambda^2} \partial_{\mu}  ( \phi^+ A_{\nu} \phi  ) - \frac{1}{\Lambda^2} \partial_{\mu}  \phi^+ \partial_{\nu}  \phi \big ] - \big [\mu \leftrightarrow \nu \big ] .
\eeqa

A candidate for a regularized form of the YM  action can be  defined as
\beq
\label{Sinv1}
 S(A,\phi^+,\phi)= \frac{1}{2g^2}\Tr \int dz F^{Inv ~ \mu \nu}(z) F_{\mu \nu}^{Inv}(z) ,
\eeq
$g$ being the dimensionless strong coupling constant.
The explicit form of this action can be expressed as
\beq
\label{Sinv2}
 S(A,\phi^+,\phi)= S_{YM} + S_{A \phi^+ \phi },
\eeq
where  the pure gauge part of this action is given in terms of   the regularized YM stress tensor  (\ref{Ainv4}) by the expression
\beq
\label{puregauge1}
S_{YM} = \frac{1}{2g^2}\Tr \int dx dy dz ~ F^{Reg ~\mu \nu}(z,x) F_{\mu \nu}^{Reg}(z,y),
\eeq
and where $S_{A\phi^+ \phi}$ is  given in terms of  the operators $G_{\mu \nu}$  (\ref{G1}), $H_{\mu \nu}$  (\ref{H1}) and  $ F_{\mu \nu}^{Reg}$  (\ref{Ainv4}) by,
\beqa
\label{Sinv3}
  S_{A\phi^+ \phi}&=& \frac{1}{2g^2}\Tr \int dx dy dz ~\bigg \{ \rho_2 (z,x) \rho_2 (z,y)G^{\mu \nu}G_{\mu \nu}(y) \nonumber \\
&& +  (\rho^{(-1)}_1 (z,x ) -\rho_2 (z,x))(\rho^{(-1)}_1 (z,y ) -\rho_2 (z,y))H^{\mu \nu}H_{\mu \nu}(y) \nonumber \\
&&  +2 \Big [ \rho_2 (z,x)G^{\mu \nu}  F_{\mu \nu}^{Reg}(z,y) +  (\rho^{(-1)}_1 (z,x ) -\rho_2 (z,x))H^{\mu \nu} F_{\mu \nu}^{Reg}(z,y) \nonumber \\
&& + \rho_2 (z,x)(\rho^{(-1)}_1 (z,y ) -\rho_2 (z,y))G^{\mu \nu}H_{\mu \nu}(y)\Big ] \bigg \}.
\eeqa

Under what conditions is the action (\ref{Sinv2}) regularized ?
In order to study this point, we rewrite the action  (\ref{Sinv3}) as a sum of five terms 
\beq
\label{Sinv4}
  S_{A\phi^+ \phi}=S_{GG} + S_{HH}  +S_{GF} +S_{HF} + S_{GH},
\eeq
each term being respectively associated to the product of the operators which enter   the expression of the  action (\ref{Sinv3}) from the left to the right.
The   Fourier transform of these terms  are given in  Appendix C.
First of all, we suppose that the cutoff functions $\rho_1$ and  $\rho_2$  (\ref{cutf1}) are chosen in such a way that the product $\rho_1^{-1}\rho_2$ are  still rapidly decreasing functions of $k^2$ in momentum Euclidean space.
The simplest choice is
\beq
\label{cutofphi0}
\rho_2(k) \sim \rho_1^2(k).
\eeq
Then, since the kinetic term arising from the term $  F_{\mu \nu}^{Reg}$ (\ref{puregauge1}) and (\ref{Ainv4})  generates a propagator  \footnote{In the sequel, we suppose that  some gauge fixing mechanism will be implemented in order that the kernel of  the kinetic term can be inverted.} proportional to $\rho_1^2$  in momentum space,  all the terms of the action  (\ref{Sinv2})  are  regularized with respect to the gauge field, except the terms of (\ref{Sinv3}) which do not contain the factor $\rho_2 $, i.e. the terms  contained in the expressions  $S_{HH}$ and $S_{HF}$  (\ref{Sinv4}).
The divergent function $\rho_1^{-2}$ enters in both the Fourier transform of these  expressions, which are  given respectively in  Appendix C by  (\ref{term1four}) and   (\ref{term2four}).
In order to regularize  them, we assume that the field $ \phi$ is in fact a smeared field defined by
\beq
\label{smearphi1}
\phi =\int dz ~\rho_3 (x,z) \varphi (z).
\eeq 
In addition, one requires, 1) that the  function $\rho_3$, which enters in the definition  (\ref{smearphi1}) of the smeared field $\phi$, behaves like 
\beq
\label{cutofphi1}
\rho_3(k) \sim \rho_1^2(k),
\eeq
as the cutoff $\Lambda$ goes to infinity, and 2) that the divergent cutoff function $\rho_1^{-1}(k)$ occurring in the expressions (\ref{term1four}) and   (\ref{term2four}), must be  multiplied by the larger number of the  smeared field $\phi$ which is a  function  of the variable $k$. 
This means that, in order to calculate the Fourier transform of the  expressions $S_{HH}$ and $S_{HF}$, the Dirac delta function that  reflects the momentum conservation at each vertex, must be   integrated over the momentum variable of the   $\phi$ field first.
In that case, as explained in Appendix C, the shift of the momentum variable of the field $\phi$ is in general not allowed in some terms of the expressions (\ref{term1four}) and   (\ref{term2four}), if these terms  become not regularized.

Keeping in mind the rules outlined before, we see that both the expressions (\ref{term1four}) and   (\ref{term2four}) are indeed regularized in  the gauge field sector, if  the propagator of the gluon field  is  a rapidly decreasing function in  Euclidean momentum space.
Moreover, given in  Appendix C, the explicit form in momentum space  of all the interaction terms between the gauge field $ A_{\mu}$ and the smeared bosons fields $\phi$ (\ref{smearphi1}),  which contribute to the action (\ref{Sinv2}) and  (\ref{Sinv4}), it is clear that  the action (\ref{Sinv2}) is now completely regularized in the gauge and scalar fields sector  as long as the propagators of the $\varphi$ fields  (\ref{smearphi1}) are also rapidly decreasing functions in  Euclidean momentum space.

Up to now the action (\ref{Sinv2}) does not contain any sector that is purely quadratic in the  scalar  field $\varphi$.
In order to define the gluon and scalar field propagators, we will add to the action  a new term, which is invariant under both the transformations  (\ref{gaugetrans0}) and  (\ref{gaugetrans2}) of the gauge and scalar fields.
This invariant term enables us to invert the quadratic part of the gauge field $ A_{\mu}$.
Since this new term  contains a quadratic part in the field $\varphi$, which does not enforce the regularization, we will also add to the action (\ref{Sinv2}) a mass term  for the fields $\varphi$ which will render the theory completely regularized.
\mysection{The scalar field sector}
Our aim is to add to the  action  (\ref{Sinv2}) a new term in order that the gluon kinetic term can be inverted without  explicitly breaking the  invariance under both the gauge  transformations  (\ref{gaugetrans0}) and (\ref{gaugetrans2}) .
The new action, thus constructed, must be gauge-invariant in all the range of the cutoff $\Lambda$, regularized and must  converge to the standard one of QCD as the cutoff goes to infinity.
Because the part (\ref{Sinv2}) of this  new action is already converging to the standard form of QCD as the cutoff goes to infinity, the searched new term must vanish  as the cutoff goes to infinity.
Then, the part of the action which will give rise to the  covariant form of the  gauge field propagator can be defined as
\beq
\label{Sgauge1}
S_{Gauge}=  \frac{\xi}{g^2} \Tr \int dx dy dz ~(\rho^{(-1)}_1 (z,x ) -\rho_2 (z,x))(\rho^{(-1)}_1 (z,y ) -\rho_2 (z,y))\partial^{\mu}\mathcal{A}_{\mu}^{Inv} \partial^{\nu}_y \mathcal{A}_{\nu}^{Inv}(y),
\eeq
the gauge-invariant operator  $\mathcal{A}_{\mu}^{Inv}$ being defined in  (\ref{Ainv1}), the scalar functions $\rho^{(-1)}_1$ and $\rho_2$ being respectively defined in
(\ref{cutf0}) and (\ref{cutf1}), and $\xi$ is a real parameter.
If we recall that the operator $U$ which enters in the definition of  $\mathcal{A}_{\mu}^{Inv}$ is expressed in terms of the smeared field $\phi$ (\ref{smearphi1}) by  (\ref{Umatrix1}), we obtain in momentum space \footnote{Remember that the limit $\eta \to 0$ must be taken in the Fourier transformed of the function $\rho^{(-1)}_1$.}
\beqa
\label{Sgauge2}
 S_{Gauge}&=& \frac{\xi}{g^2} \Tr \int \dk  ~(\rho_1^{-1} (k ) -\rho_2 (k))^2 \Big [ k_{\mu}  k_{\nu}A^{\mu}(k) A^{\nu}(-k)  + \frac{ k^4}{4\Lambda^2} (\phi^+(k) - \phi(-k))\nonumber \\
&& \times (\phi^+(-k) - \phi(k))      \Big] + S_{Gauge}(A,\phi),
\eeqa
where the term $ S_{Gauge}(A,\phi)$ is explicitly given by the expression (\ref{SgaugeAphi1}).
Adding both the quadratic part  in the gluons fields of  the expression (\ref{Sgauge2}),  to that  of the regularized YM action  $S_{YM}$ (\ref{puregauge1}), allows us to invert the gauge field kinetic term, and thus to define the free gluon propagator.
This propagator is given in momentum space by
\beq
\label{propgluon1}
\langle A_{\mu \alpha}   A_{\nu \beta} \rangle  = - \rho_1^2 (k )\frac{1}{k^4} \delta_{ \alpha  \beta} \big[ (k^2g_{\mu \nu} -k_{\mu}k_{\nu} ) + \frac{1}{\xi (1 -\rho_1 (k )\rho_2 (k))^2}k_{\mu}k_{\nu} \big ].
\eeq

What happens in the pure scalar fields sector ?
We define  the smeared field $\phi$  in terms of the smeared fields  $ \sigma$ and $\pi$ by
\beq
\label{phi2}
\phi = \sigma + i\pi ,
\eeq
and, in the same way that  the smeared field $\phi$   (\ref{Umatrix2}) is associated to the field $\varphi$  in  (\ref{smearphi1}), we define  the anti-Hermitic matrices   $ \varsigma$ and $\varpi$ as
\beq
\label{phi1}
\varphi = \varsigma + i\varpi,
\eeq
whose components  $\varsigma_i$ and $\varpi_i$, on the basis formed by $T_0$ and the eight generators   $T_{\alpha}$, are real scalar fields.
Because  the fields $\varsigma_a$ and $\varpi_a$ are real, we have,
\beqa
\label{realchi1}
\varsigma^+(k)&=&-\varsigma(-k) \\
\label{realchi2}
\varpi^+(k)&=&-\varpi(-k).
\eeqa
In that case,  the second term of (\ref{Sgauge2}) will give only a contribution to the propagator of the $\varsigma$  field.
Up to now, since there is no other  quadratic term in the pure scalar fields sectors of the parts  of the action given in  (\ref{Sinv2}) and   (\ref{Sgauge2}), and knowing that the cutoff function $\rho_3$ which enters in the definition of the smeared  $ \sigma$ field  (\ref{phi2}) behaves like (\ref{cutofphi1}), we find that the free propagator of the  $\varsigma$  field  (\ref{phi1}) will behave respectively in the IR  and UV domains as
\beq
\label{propsigma1}
\langle \varsigma  \varsigma \rangle  \sim \frac{\Lambda^2}{k^4} \rho_1^{-2} (k ).
\eeq
Such a  behavior for the $\varsigma$ field will then spoil the UV regularization of the theory and worse will enforce the IR divergences.
Moreover a perturbative calculation cannot be performed because in the actions  (\ref{Sinv2}) and   (\ref{Sgauge2}) there is no quadratic term in the $ \varpi$ fields  (\ref{phi1}).
To cure this disease, we will add to the action a mass term for both  $ \varsigma$ and $\varpi$ fields, while preserving if possible the  invariance of the action under both the  transformations (\ref{gaugetrans0}) and (\ref{gaugetrans2}).

We will first construct a mass invariant term for the field $\varpi$ which  is proportional to the sum $\phi^+ + \phi$.
Since the operator 
\beq
\label{massechi1}
U^+U -I = \frac{1}{\Lambda} (\phi^+ +\phi) +\frac{1}{\Lambda^2}\phi^+ \phi,
\eeq
where $U$ is defined by  (\ref{Umatrix1}), is invariant under the transformation (\ref{gaugetrans2}), the searched mass term will arise from the following action
\beq
\label{massechi2}
S_{\pi} = -\frac{M^2}{4g^2}\Tr  \int dx dy  ~ \rho^{(-6)}_1 (x,y )(\phi^+ +\phi +\frac{1}{\Lambda}\phi^+ \phi)(\phi^+(y) +\phi(y) +\frac{1}{\Lambda}\phi^+(y) \phi(y)) ,
\eeq
where the function $ \rho^{(-6)}_1$ is defined in   (\ref{cutf0}) and $M$ is a mass parameter.
Therefore if we use the definitions (\ref{smearphi1}) and (\ref{phi1}), and the relations   (\ref{realchi1}) and  (\ref{realchi2}), and suppose for simplicity that the cutoff function $\rho_3$ (\ref{cutofphi1}), which enters in the definition of the smeared fields  $\sigma$ and $\pi$  (\ref{phi2}), is in fact given by
\beq
\label{cutofphi2}
\rho_3(k)= \rho_1^2(k),
\eeq
the action (\ref{massechi2}) becomes in  momentum space
\beqa
\label{massechi3}
S_{\pi}&=& \frac{M^2}{g^2}\Tr \bigg \{\int \dk~\rho_1^{-2}(k)\varpi(k) \varpi(-k) - \frac{i}{\Lambda}  \int \dk\ddp ~ \varphi^+(p) \varphi(p+k ) \varpi(-k) \nonumber \\
&& -\frac{1}{4\Lambda^2}    \int \dk\ddp \dq   ~\rho_1^2 (k )\varphi^+(p) \varphi(p-k ) \varphi^+(q) \varphi(q+k ) \bigg \}.
\eeqa
In order that the second term of (\ref{massechi3})  gives rise  only to regularized contributions, the free propagators of the   $\varsigma$ and $\varpi$   fields must be rapidly decreasing functions in Euclidean momentum space.

Regarding  the  $\varsigma$  field, there exists a priori no mass term invariant under the   transformation (\ref{gaugetrans2}).
For simplicity, we choose the following term,
\beq
\label{massesigma1}
S_{\sigma} = \frac{M^2}{4g^2}\Tr  \int dx dy  ~ \rho^{(-6)}_1 (x,y )(\phi^+ -\phi )(\phi^+(y) -\phi(y) ) ,
\eeq
for the  $\varsigma$  field mass term, which is given in momentum space by
\beq
\label{massesigma2}
S_{\sigma} =   \frac{M^2}{g^2}\Tr  \int \dk~\rho_1^{-2}(k)\varsigma(k) \varsigma(-k) .
\eeq
We will see in the following, that the breaking induced by the term  (\ref{massesigma2}) does not really spoils the gauge invariance of the theory.
Now from the relations (\ref{Sgauge2}) and (\ref{massesigma2}), we obtain for the free propagator of the  $\varsigma$ field
\beq
\label{propsigma3}
\langle \varsigma  \varsigma \rangle  =-  \frac{1}{\xi \frac{k^4}{\Lambda^2} \rho_1^2(k)(1-\rho_1(k) \rho_2(k))^2 +M^2 \rho_1^{-2}(k)}.
\eeq
From the quadratic term of the action  (\ref{massechi3}), and from the relation  (\ref{propsigma3}), we conclude that the free  propagators of the $\varpi$ and  $\varsigma$ fields have the requested behavior in the IR and UV domain, i.e.
\beq
\label{propsigma4}
\langle \varpi  \varpi \rangle  \sim \langle \varsigma  \varsigma \rangle  \sim \frac{ \rho_1^2 (k )}{M^2}.
\eeq
\mysection{The regularized gauge-invariant path integral}
From the preceding sections, we conclude that the form of the regularized action of the YM fields is given in adding to the action  (\ref{Sinv2}) both the contributions (\ref{Sgauge1}) of the gauge sector and that  of the mass terms  (\ref{massechi2}) and  (\ref{massesigma1}) of the  $\varpi$ and $\varsigma$ fields.
Then the regularized action is 
\beq
\label{fullaction1}
S_{Reg}(A,\varphi^+, \varphi,\Lambda)= S_{Inv}(A,\varphi^+, \varphi,\Lambda)+S_{\sigma},
\eeq
where 
\beq
\label{fullactioninv1}
S_{Inv}(A,\varphi^+, \varphi,\Lambda)= S_{YM} + S_{A \phi^+ \phi } +S_{Gauge} +S_{\pi},
\eeq
is invariant under the gauge transformations (\ref{gaugetrans0}) and (\ref{gaugetrans1}), and  we recall that $\Lambda$ is the cutoff scale and $M$ is the mass of the   $\varpi$ and  $\varsigma$ fields.
In the configuration space the regularized  path integral which is normalized to unity reads
\beq
\label{Zreg1}
\ Z_{Reg}(J,\Lambda)=\frac{\int \D A_{\mu}\D \varphi^+ \D \varphi ~e^{i\left(S_{Reg}(A,\varphi^+, \varphi,\Lambda) +S(J)\right)}} {\int \D A_{\mu}\D \varphi^+ \D \varphi ~e^{iS_{Reg}(A,\varphi^+, \varphi,\Lambda)} },
\eeq
where $S(J)$ is given as usual in terms  of the external sources  $J_{\mu}$ for the gluons fields by 
\beq
\label{source1}
S(J)= -\frac{2}{g}\Tr \int dx~J_{\mu }A^{\mu }.
\eeq
By construction, when the cutoff $\Lambda$ goes to infinity, the action (\ref{fullaction1}) has the property
\beq
\label{fullaction2}
\lim_{\Lambda \to \infty} S_{Reg}(A,\varphi^+, \varphi,\Lambda)= \frac{1}{2g^2}\Tr \int dx F^{ \mu \nu} F_{\mu \nu}  - \frac{M^2}{g^2}\Tr \int dx  \varphi^+ \varphi.
\eeq
This tells us that the auxiliary bosons fields $ \varphi$ will decouple from the theory.
We will show that this decoupling occurs when the mass $M$ of the  bosons fields $ \varphi$ are at least greater or equal to the cutoff scale $\Lambda$.

After a loopwise expansion of the path integral (\ref{Zreg1}) with respect to the part of the action which contains the vertices of the  $ \varphi$  fields and integration over these fields, as show in  Appendix D, the proper vertex $\Gamma$ (\ref{vertex5}) associated to any  gluons effective amplitude behaves like
\beq
\label{decouple1}
\Gamma  \sim  \big  (\frac{ \Lambda^2}{M^2}\big  )^{I_{\varphi}-n} \mathcal {O} (\Lambda ^{-k}),
\eeq
as all the internal momenta scale with the cutoff  $\Lambda$.
Here $n$ and $I_{\varphi}$ are respectively  the number of pure  $ \varphi$  fields vertices and the number of internal  $ \varphi$  fields occurring in $\Gamma$.
As show in  Appendix D, the number   $k$  (\ref{topology4}) is a positive integer for all the gluons effective  vertices, except for the two, three and four legs effective vertices, which are proportional to the vertices of the pure Yang-Mills action (\ref{puregauge2}), where  $k$ is zero.
As a result if 
\beq
\label{decouple2}
1 \ll \Lambda^2 \leq M^2 \quad \text{or} \quad   0 \leq \Lambda^2 \ll M^2,
\eeq
 the  auxiliary  fields  $\varphi$ decouple from the gauge sector.
Notice that the  rescaling
\beq
\label {resaling1}
 \phi~ \rightarrow  ~  \Lambda  \phi,
\eeq
of the smeared field  $\phi$  (\ref{smearphi1}) does not change this conclusion.
This is because, in each internal $\varphi$ boson line the rescaling of the field is exactly compensated by the rescaling 
\beq
\label {resaling2}
 M~ \rightarrow  ~  \Lambda  M,
\eeq
of the mass $M$ entering in the  propagator, so that the behavior of the proper effective vertex $\Gamma$ (\ref{decouple1}) is left unchanged.
In order to have only one scale in the  regularized  path integral (\ref{Zreg1}), a good choice is to take for the mass $M$ of the  $\varphi$  fields the   cutoff scale $\Lambda$.

If we denote $\rho_1$ by $\rho$, and remember that the cutoff function  $\rho_2$ is constrained by the relation  (\ref{cutofphi0}), and that we have fixed the function $\rho_3$ by the equation  (\ref{cutofphi2}), the action  (\ref{fullaction1}) is still regularized if we choose the following  relations between $\rho_2$ and  $\rho_3$
\beq
\label{newcutoff2}
\rho_2(k)=\rho_3(k)=\rho(k)^2.
\eeq

What are the WT identities that  can be deduced from the expression of the path integral  (\ref{Zreg1}) ?
The  WT identities follow from the invariance of the  path integral (\ref{Zreg1}) under the shift of variable induces by the  infinitesimal  form of the gauge transformation  (\ref{gaugetrans0}) and the smeared field transformation  (\ref{gaugetrans2}), i.e.
\beqa
\label{gaugetrans3}
 A^{\mu}_{\alpha} ~&\rightarrow &~ A^{\mu}_{\alpha} -C_{\alpha \beta \gamma}A^{\mu}_{\beta} \epsilon_{ \gamma} +\partial_{\mu}\epsilon_{\alpha} \\
\label{gaugetrans4}
\phi  ~&\rightarrow &~ \phi +\epsilon \phi +\Lambda\epsilon,
\eeqa
where we use the shorthand notation $\epsilon \equiv \epsilon_{\alpha}T^{\alpha}$.
By construction, all the terms of the action (\ref{fullaction1}), except the term $S_{\sigma}$, are invariant under both the gauge transformation  (\ref{gaugetrans0}) and the smeared field transformation  (\ref{gaugetrans2}).
In addition the  measure $ \D \varphi^+ \D \varphi$ is also invariant under the transformations of the fields  $\varphi$ which are induced by the transformations  (\ref{gaugetrans2})    of the smeared field $\phi$ (\ref{smearphi1}).
This is because, 1) the measure $ \D \varphi^+ \D \varphi$  is proportional\footnote{This fact is readily seen in momentum space.}  to  the measure $\D \phi^+ \D \phi$ and 2)  the transformation  (\ref{gaugetrans2}) of the smeared field $\phi$  (\ref{smearphi1}) leaves the measure   (\ref{scalarmeasure1}) invariant.

In order to derive the WT identities for the gluons fields, we will show that the symmetry breaking term  $S_{\sigma}$  (\ref{massesigma1}) is harmless.
In fact, if we introduce the auxiliary matrix valued field 
\beq
\label{varrho1}
\varrho=\varrho_aT^a,
\eeq
whose   components $\varrho_a$ are real scalar fields, and define the action 
\beq
\label{actionvarrho1}
S_{\sigma \varrho}= - \frac{M^2}{g^2}\Tr \int dy \big [2  \varrho(y)  \int dx \rho^{(-1)}(x,y) \sigma +  \varrho^2(y) \big ],
\eeq
the path integral (\ref{Zreg1}) can be written as
\beq
\label{Zreg2}
\ Z_{Reg}(J,\Lambda)=\frac{\int \D A_{\mu}\D \varphi^+ \D \varphi  \D \varrho ~e^{i\left(S_{Inv}(A,\varphi^+, \varphi,\Lambda) + S_{\sigma \varrho} +S(J)\right)}} {\int \D A_{\mu}\D \varphi^+ \D \varphi  \D \varrho~e^{i \left(S_{Inv}(A,\varphi^+, \varphi,\Lambda)+S_{\sigma \varrho}\right ) } }.
\eeq
This is because the Gaussian integration over the  $\varrho$ fields transforms the action $S_{\sigma  \varrho}$ (\ref{actionvarrho1}) in the action $S_{\sigma}$ (\ref{massesigma2}).
Since under the infinitesimal transformation (\ref{gaugetrans4}) the components of the smeared fields $\sigma$ and $\pi$ are mixed in the following way
\beqa
\label{gaugetrans5}
\delta \sigma_0& =&- \frac{1}{\sqrt{6}} \pi_{\alpha}\epsilon_{\alpha}  \\
\delta \sigma_{\alpha}& =& -\frac{1}{2} (C_{\alpha \beta \gamma}\sigma_{\beta}+d_{\alpha \beta \gamma}\pi_{\beta})\epsilon_{\gamma} - \frac{1}{\sqrt{6}}\pi_0\epsilon_{\alpha} +\Lambda\epsilon_{\alpha} \nonumber \\
\label{gaugetrans6}
\delta \pi_0& =& \frac{1}{\sqrt{6}} \sigma_{\alpha}\epsilon_{\alpha} \\
\delta \pi_{\alpha}& =& -\frac{1}{2} (C_{\alpha \beta \gamma}\pi_{\beta} -d_{\alpha \beta \gamma}\sigma_{\beta})\epsilon_{\gamma} + \frac{1}{\sqrt{6}}\sigma_0\epsilon_{\alpha}  \nonumber,
\eeqa
the infinitesimal variation of the action  $S_{\sigma  \varrho}$ (\ref{actionvarrho1}) under the transformation (\ref{gaugetrans4}) is given by
\beqa
\label{actionvarrho2}
\delta S_{\sigma \varrho}&=&- \frac{M^2}{g^2} \int dx dy  \rho^{(-1)}(x,y) \Big [\frac{1}{\sqrt{6}}  \varrho_0(y) \epsilon_{\alpha}\pi_{\alpha} +  \varrho_{\beta}(y)\big \{\frac{1}{2} \epsilon_{\alpha} ( \sigma_{\gamma}C_{\alpha \beta \gamma} +\pi_{\gamma}d_{\alpha \beta \gamma}) \nonumber \\
&&+ (\frac{1}{\sqrt{6}} \pi_0 -\Lambda) \epsilon_{\beta} \big\} \Big ].
\eeqa
The invariance under the infinitesimal transformations (\ref{gaugetrans3}), (\ref{gaugetrans5}) and (\ref{gaugetrans6}) of the path integral (\ref{Zreg2}) gives the following identity
\beq
\label{WT1}
\bra \frac{\delta S_{\sigma \varrho}}{\delta \epsilon_{\alpha}} - \frac{1}{g} (C_{\alpha \beta \gamma}J^{\mu}_{\beta}  A_{\mu \gamma} +\partial_{\mu}J^{\mu}_{\alpha} ) \ket =0. 
\eeq
In a intermediate step, and in order to express the   connected Green's functions,  we introduce  respectively   in the path integral  (\ref{Zreg2}) the external sources $K_a$, $L_a$ and $N_a$ for the scalar fields $\varsigma_a$, $\varpi_a$  (\ref{phi1}) and $\varrho_a$  (\ref{varrho1}), in replacing the source term $S(J)$ (\ref{source1}) by the expression
\beq
\label{source2} 
S(J,K,L,N)= -\frac{2}{g}\Tr \int dx~(J_{\mu }A^{\mu }+ K\varsigma + L\varpi + N\varrho).
\eeq
Then, the  generating functional of connected Green's functions in the presence of the external sources for the gauge and scalar fields is defined as
\beq
\label{Zreg3}
\ Z_{Reg}(J,K,L,N,\Lambda)=e^{W_{Reg}(J,K,L,N,\Lambda)}.
\eeq
As usual,  we define for each field its  vacuum  expectation value in the presence of the external source.
These C-number functions are  obtained by taking the functional derivative of $W_{Reg}$ with respect to their  own sources.
We also recall that the generating functional of 1PI functions is defined in terms of the  vacuum  expectation values of the fields by the Legendre transformation
\beq
\label{1PIgene1}
\Gamma_{Reg}(A, \varsigma,\varpi,\varrho,\Lambda)=-iW_{Reg}(J,K,L,N,\Lambda) + \frac{2}{g}\Tr \int dx~(J_{\mu }A^{\mu }+ K\varsigma + L\varpi + N\varrho).
\eeq
In that case there is a one-to-one correspondence between the classical field thus obtained and its external source.
For instance these    conjugate relations hold true for  the  fields  $A_{\mu}$ and  $\varrho$ 
\begin{alignat}{3}
\label{vacuumexp1}
A^{\mu}_{\alpha} &=&  -ig\frac{\delta W_{Reg}}{\delta J_{\mu \alpha}}& \qquad J_{\mu \alpha} &=&  -g\frac{\delta \Gamma_{Reg}}{\delta A^{\mu}_{ \alpha}} \\
\label{vacuumexp2}
\varrho_a &=&  -ig\frac{\delta W_{Reg}}{\delta N_a} & \qquad N_a &=&  -g\frac{\delta \Gamma_{Reg}}{\delta  \varrho_a}  
\end{alignat}

We are now able to  show that the functional derivative of $\bra \frac{\delta S_{\sigma \varrho}}{\delta \epsilon_{\alpha}} \ket$ (\ref{actionvarrho2}) with respect to the  vacuum expectation value of the gauge field $A_{\mu}$ vanishes identically.

At first, when taking  the  functional derivative of $\bra \frac{\delta S_{\sigma \varrho}}{\delta \epsilon_{\alpha}} \ket$ (\ref{actionvarrho2}) with respect to the classical field  $A_{\mu}$, the terms which are only proportional to $\bra \varrho \ket$ do not contribute.
This is because,  1) the vacuum  expectation values  of the  fields  $A_{\mu}$ and $\varrho$  are by construction independent variables, and 2)  at the end of the calculation we will set  to zero the external sources relative to the scalar fields,
Then, using the chain rule which expresses the functional derivative $\frac{\delta}{\delta A}$ in term of  $\frac{\delta}{\delta J}$, it remains to show that  the functional derivative of the connected Green's functions $\bra \varrho_a \pi_b \ket$ and $\bra \varrho_a \sigma_b \ket$ with respect to the gauge fields sources $J_{\mu}$ vanish effectively.

The equation of motion of the auxiliary field $\varrho$, which is deduced  from the expressions of the  generating functional (\ref{Zreg3}) and  of the action $S_{\sigma \varrho}$ (\ref{actionvarrho1}), can be expressed with the help of the relations  (\ref{vacuumexp2}) in terms of the classical  fields as
\beq
\label{equamotion1}
 M^2 [ \int dy ~ \rho(x,y)  \varsigma_a(y) +\varrho_a ] -g^2  \frac{\delta \Gamma_{Reg}}{\delta  \varrho_a}  =0.
\eeq
If   we take  the functional derivative of both sides  of the equation  (\ref{equamotion1}), first  with respect to the classical fields $ \varpi_b(y)$  or $ \varsigma_b(y)$, and then with respect to  $A^{\mu}_{\alpha}(z)$, we get 
\beqa
\label{equamotion2}
 \frac{\delta^3 \Gamma_{Reg}}{  \delta A^{\mu}_{\alpha} (z) \delta \varpi_b(y) \delta \varrho_a }& =& 0 \nonumber \\
  \frac{\delta^3 \Gamma_{Reg}}{  \delta A^{\mu}_{\alpha} (z) \delta \varsigma_b(y) \delta \varrho_a }& =& 0. 
\eeqa
Because $\pi$ and $\sigma$ are respectively  the smeared field associated to the scalar field  $ \varpi$ and $ \varsigma$, the vanishing of the 1PI functions  (\ref{equamotion2})  shows that the functional derivatives of the connected Green's functions  $ \bra \varrho_a \pi_b \ket$ and $ \bra \varrho_a \sigma_b \ket$ with respect to the source $J$ of the gauge field also vanish identically, whatever the values of the scalar fields  sources.
As a result  the functional derivative  of the vacuum expectation value $\bra \frac{\delta S_{\sigma \varrho}}{\delta \epsilon_{\alpha}} \ket$ (\ref{actionvarrho2}) with respect to the classical  field $A_{\mu}$, indeed vanishes.
Then,  in expressing  the WT identity (\ref{WT1}) in terms of 1PI functions, we get the following relation
\beq
\label{WT2}
 \frac{\delta^2 \Gamma_{Reg}}{ \delta  A^{\nu}_{\beta} (y) \delta A^{\mu}_{\gamma} }C_{\alpha \gamma \delta} A^{\mu}_{\delta} -C_{\alpha \beta \gamma}\frac{\delta \Gamma_{Reg}}{  \delta A^{\nu}_{\gamma}} \delta (x-y) + \partial^{\mu} \frac{\delta^2 \Gamma_{Reg}}{  \delta A^{\mu}_{\alpha} \delta  A^{\nu}_{\beta} (y)} =0.
\eeq
In this identity $ \Gamma_{Reg}$ is the  generating functional of 1PI functions defined by
\beq
\label{1PIgene2}
\Gamma_{Reg}(A, \Lambda)=-iW_{Reg}(J,\Lambda) + \frac{2}{g}\Tr \int dx~J_{\mu }A^{\mu },
\eeq
which means that the integration over the scalar fields $\varphi$ has already been done, at least implicitly, in the path integral  (\ref{Zreg1}).
The regularized WT (\ref{WT2}) is our main result, and shows that  all regularized 1PI functions which are deduced from the regularized  path integral  (\ref{Zreg1})
 are all transverse with respect to the gauge field when the external source is switched off.
\mysection{Conclusion and outlook} 
We have shown that the path integral of pure YM theory can be regularized in four-dimensional physical space in a gauge-invariant manner.
The regularization is implemented nonperturbatively at the level of the action through cutoff functions.
The price to pay, in order to maintain the original gauge invariance, is to enlarge the field content of the theory by a set of auxiliary complex scalar fields that play the role of Stueckelberg compensating fields.
By contrast with any PV inspired regularization, the Gaussian metric induced by  these fields has the good sign, and the path integral is here a mathematically well-defined object. 
Since the action, and therefore the  path integral of the theory, is regularized nonperturbatively in all the energy range, this regularization will be useful to study the evolution of physical quantities with  the energy scale, without making any assumptions for matching of the quantities calculated in the IR and in the UV domains, as  is the case in lattice gauge theory.
In a forthcoming paper, we shall study  first the consequences  of this regularization in perturbation theory, and then beyond.
\vskip 5mm
\centerline{\bf Acknowledgements}
I am grateful to  J. Polonyi, M. Rausch de Traubenberg and M.J. Slupinski for useful and enlightening discussions, and to N. Rivier for the careful reading of the manuscript.
\vskip 8mm
\myappendix{Appendix A } 
\appA
Here  for convenience and to fix our conventions we recall the commutation relations  of the generators (\ref{generator1}) of the group $SU(3)$ 
\beqa
\label{Tcom1}
[T^{\alpha},T^{\beta}]&=&C_{\alpha \beta \gamma}T^{ \gamma} \\
\label{Tcom2}
\{T^{\alpha},T^{\beta}\}&=&-\frac{1}{3} \delta_{\alpha \beta} +id_{\alpha \beta \gamma}T^{ \gamma}, 
\eeqa
the  structures constants $C_{\alpha \beta \gamma} $ having the opposite sign of those defined in \cite{ZUBER}. 
We recall that the trace properties of these matrices follow recursively  from the relation 
\beq
\label{Tcom0}
T^{\alpha}T^{\beta}=-\frac{1}{6} \delta_{\alpha \beta} + \frac{1}{2}(C_{\alpha \beta \gamma} + id_{\alpha \beta \gamma})T^{ \gamma}.
\eeq
For instance the trace of the product of three and four generators are given by 
\beqa
\label{Tcom3}
\Tr( T^{\alpha}T^{\beta} T^{ \gamma})&=& -\frac{1}{4}( C_{\alpha \beta \gamma} +i d_{\alpha \beta \gamma} ) \\
\label{Tcom4}
\Tr( T^{\alpha}T^{\beta} T^{ \gamma} T^{\delta}) &=&\frac{1}{12} \delta_{\alpha \beta}  \delta_{ \gamma \delta } -\frac{1}{8}( C_{\alpha \beta \alpha'} +i d_{\alpha \beta \alpha'} ) ( C_{\gamma \delta \alpha'} +i d_{\gamma \delta \alpha'} ) .
\eeqa
\myappendix{Appendix B } 
\appB
To have an explicit expression of the transformation  of the scalar fields $\phi_i$ induced by the matrix transformation (\ref{gaugetrans2}), we  recast this  equality between two $3 \times 3$  complex  matrices as an equality between two complex vectors having nine components.
Let $C_V$ be  the  nine components vector column obtained from  any  $3 \times 3$  matrix $V$ by putting the successive columns of the matrix $V$ as a single column in increasing order from top to bottom.
Then,  the tensorial product involving  the complex $3 \times 3$  matrix $V$ 
\beq
\label{Tmatrix0}
T(V)=I \otimes V,
\eeq
is a $9 \times 9$ matrix $T(V)$ that relates the constant  column  vector $C_I$  associated with the  $3 \times 3$ identity matrix  $I$ to the  column  vector $C_V$ as,
\beq
\label{Tmatrix1}
C_V=T(V) C_I.
\eeq
By construction, the linear mapping $T$  between the set of $3 \times 3$ complex matrices $V$ and the set of  $9 \times 9$ complex matrices $T(V)$, is an group isomorphism, that is to say, for any matrices $V$ and $V'$  the relations 
\begin{align}
\label{Tmatrix2}
T(VV')=T(V)T(V'), \quad T(I)=I  \quad \text{and} \quad T(V^+)=T^+(V)
\end{align}
hold.
Moreover, due to the definition (\ref{Tmatrix0}) we have the obvious relation 
\begin{align}
\label{Tmatrix3}
\det T(V) = (\det V )^3.
\end{align}
Using the definition  (\ref{Tmatrix1}) and the properties  (\ref{Tmatrix2}), the vectorial form of the equation (\ref{gaugetrans2}) is then given by the equation
\begin{align}
\label{Tmatrix4}
T( \phi')C_I=T(V)T( \phi)C_I +\Lambda T(V-I)C_I  .
\end{align}
We can write the vector $T( \phi)C_I$  associated to the matrix (\ref{Umatrix2}) as
\begin{align}
\label{Tmatrix5}
T( \phi)C_I=S \Phi,
\end{align}
where $S$ is an invertible $9 \times 9$ constant matrix acting on the vector  column $ \Phi$, whose components are the fields $ \phi_i$ defined in  (\ref{Umatrix2}), the index $i$ running from $0$ to $8$.
In that case, we can rewrite  the equation (\ref{Tmatrix4}) as
\begin{align}
\label{Tmatrix7}
S \Phi'=T(V)S \Phi +\Lambda T(V-I)C_I .
\end{align}
\myappendix{Appendix C } 
\appC
In this appendix we give explicitly the terms of the action  (\ref{Sinv2}) in momentum space.
In all the terms,  the integration over the Dirac's delta functions which result from the  translational invariance of the action are  first performed with respect to  the momenta of the $\phi$ field, and owing to the conditions (\ref{cutofphi0}) and  (\ref{cutofphi1}) the product of cutoff functions $\rho_1^{-1}\rho_2$ and  $\rho_1^{-1}\rho_3$ are rapidly decreasing functions of the square of the momenta in Euclidean space.
Moreover, as said before, the inverse Fourier transform of the  scalar function (\ref{cutf0}) $\rho^{(-1)}_1$ is defined when the infinitesimal parameter $\eta$ tends to zero. 
After a lengthy but straightforward calculation, we obtain respectively for the terms $S_{HH}$,  $S_{HF}$ and  $S_{Gauge}$ which are defined in (\ref{Sinv3}) , (\ref{Sinv4}) and (\ref{Sgauge2}).
\beqa
\label{term1four}
S_{HH}&=& \frac{1}{g^2} \Tr  \int \dk \ddp \dq~ (\rho_1^{-1} (k)-\rho_2 (k))^2 \bigg  \{   (k^2q_{\mu }-kqk_{\mu}) A^{\mu}_{\alpha} (p) \Big [ \frac{2i}{\Lambda^3} (\phi^+(p-k)T^{\alpha} \nonumber \\
&& + T^{\alpha}\phi(k-p))   \phi^+(q )\phi(q-k)   + \frac{2i}{\Lambda^4} \int \ddp' \phi^+(p')T^{\alpha}\phi(p'+k-p)\phi^+(q )\phi(q-k) \Big ]  \nonumber \\
&& +  (k^2g_{\mu \nu}-k_{\mu}k_{\nu}) A^{\mu}_{\alpha} (p)  A^{\nu}_{\beta} (q) \Big [ \frac{1}{\Lambda^2} (\phi^+(p-k)T^{\alpha}     +T^{\alpha} \phi(k-p))   (  \phi^+(q +k)T^{\beta}\nonumber \\
&&  +T^{\beta}\phi(-q -k))   + \frac{2}{\Lambda^3}   \int  \ddp' ( \phi^+(p-k)T^{\alpha}   +  T^{\alpha}\phi(k-p))  \phi^+(p') T^{\beta} \phi(p'-k-q) \nonumber \\
&&   +\frac{1}{\Lambda^4} \int \ddp' \dq' \phi^+(p')T^{\alpha} \phi(p'+k-p)  \phi^+(q')T^{\beta}  \phi(q'-k-q)\Big ]  \nonumber \\
&&  -\frac{1}{\Lambda^4} p^{\mu} (k^2q_{\mu }-kqk_{\mu})\phi^+(p)\phi(k+p) \phi^+(q)\phi(q -k) \bigg \}
\eeqa 
 \beqa
\label{term2four}
S_{HF}&=&  \frac{2}{g^2} \Tr \int  \ddp \dq ~\rho_1^{-1} (p)(\rho_1^{-1} (p)-\rho_2 (p))  \bigg  \{  \frac{i}{\Lambda^2}    (p^2q_{\mu }-pqp_{\mu}) A^{\mu}_{\alpha} (p) T^{\alpha}\phi^+(q )  \phi(q-p )   \nonumber \\
&& +  (p^2g_{\mu \nu}-p_{\mu}p_{\nu}) A^{\mu}_{\alpha} (p)  A^{\nu}_{\beta} (q)  \Big [ \frac{1}{\Lambda} (T^{\alpha}T^{\beta}\phi(-p-q )   + T^{\beta}T^{\alpha}\phi^+(p+q ) )  \nonumber \\
&& +  \frac{1}{\Lambda^2}  \int  \ddp'  T^{\alpha}\phi^+(p' )T^{\beta} \phi(p' -p-q )   \Big ]  \bigg \}  -  \frac{2}{g^2} \Tr \int  \ddp \dq \dr \rho_2 (q+r)(\rho_1^{-1} (q+r) \nonumber \\
&& -\rho_2 (q+r))   \bigg  \{  \frac{1}{\Lambda^2} p_{\nu} (q+r)_{\mu} A^{\nu}_{\beta} (q) A^{\mu}_{\gamma} (r)  \phi^+(p) \phi(p-q-r) [T^{\beta},T^{\gamma}] \nonumber \\
&& +(q+r)_{\nu}  A^{\mu}_{\alpha} (p)  A^{\nu}_{\beta} (q) A_{\gamma \mu} (r) \Big [
\frac{i}{\Lambda} ( \phi^+(p+q+r)T^{\alpha} + T^{\alpha} \phi(-p-q-r))  \nonumber \\
&&+ \frac{i}{\Lambda^2} \int  \ddp'  \phi^+(p')T^{\alpha} \phi(p'-p-q-r) \Big ] [T^{\beta},T^{\gamma}]   \bigg \}
\eeqa 
\beqa
\label{SgaugeAphi1}
S_{Gauge}(A,\phi)&=& \frac{\xi}{g^2} \Tr \int \dk  ~(\rho_1^{-1} (k ) -\rho_2 (k))^2  \bigg  \{   \int \ddp k_{\mu} A^{\mu}_{\alpha} (p) \Big [ \frac{i}{\Lambda^2}k^2 (\phi^+(p-k)T^{\alpha}   \nonumber \\
&& +T^{\alpha} \phi(k-p))  (  \phi^+(k)-\phi( -k))  +  \frac{i}{\Lambda^3}  \int \dq k^2  \phi^+(q)T^{\alpha}\phi(q+k-p) \nonumber \\
&& \times  (  \phi^+(k)-\phi( -k)) + \frac{i}{\Lambda^3}  \int  \dq (q^2 -(k-q)^2) (\phi^+(p-k)T^{\alpha}  \nonumber \\
&&+ T^{\alpha}\phi(k-p))  \phi^+(q) \phi(q-k) +  \frac{i}{\Lambda^4}  \int \ddp' \dq (q^2 -(k-q)^2) \phi^+(p')T^{\alpha} \nonumber \\
&& \times  \phi(p'+k-p)  \phi^+(q)\phi(q -k)  \Big ]  +  \int \ddp  \dq  k_{\mu}  k_{\nu}A^{\mu}_{\alpha} (p)A^{\nu}_{\beta} (q)   \Big [ \frac{1}{\Lambda^2} (\phi^+(p-k)T^{\alpha}  \nonumber \\
&& +T^{\alpha} \phi(k-p)) ( \phi^+(k+q)T^{\beta} +T^{\beta}  \phi(-k-q)) 
 + \frac{2}{\Lambda^3}\int \ddp'~ (\phi^+(p-k)T^{\alpha}  \nonumber \\
&& +T^{\alpha} \phi(k-p))  \phi^+(p')T^{\beta}\phi(p'-k-q )
+ \frac{1}{\Lambda^4} \int \ddp' \dq'~ \phi^+(q')T^{\alpha}\phi(k-p+q')  \nonumber \\
&& \times  \phi^+(p') T^{\beta}\phi(p'-k-q )   \Big ]   + \int \ddp (p^2 -(k +p)^2)   \Big [ \frac{1}{2\Lambda^3}k^2  \phi^+(p)\phi(k+p) \nonumber \\
&& \times  (  \phi^+(k)-\phi( -k))  + \frac{1}{4\Lambda^4} \int \dq 
 (q^2 -(q-k)^2) \phi^+(p)\phi(k+p)  \phi^+(q)\phi(q -k)  \Big ]  \nonumber \\
&& +   k_{\mu} A^{\mu}_{\alpha} (k) \Big [ \frac{i}{\Lambda}k^2 T^{\alpha}(\phi^+(k ) -\phi(-k )) +  \frac{i}{\Lambda^2} \int \ddp ((k+p)^2 -p^2)T^{\alpha} \phi^+(k+p)  \nonumber \\
&& \times \phi(p) \Big ] + \int \ddp    k_{\mu}  k_{\nu}A^{\mu}_{\alpha} (k)A^{\nu}_{\beta} (p)   \Big [ \frac{2}{\Lambda}T^{\alpha} (\phi^+(k+p)T^{\beta}  +T^{\beta} \phi(-k-p))  \nonumber \\
&& +  \frac{2}{\Lambda^2} \int \dq T^{\alpha} \phi^+(q)T^{\beta}  \phi(q-k-p) \Big ]\bigg \}
\eeqa 
The divergent cutoff function $\rho_1^{-2}$ enters in both the expressions (\ref{term1four}) ,  (\ref{term2four}) and (\ref{SgaugeAphi1}).
The shift of the momentum variable of the field $\phi$ is in general not allowed in some terms of the expressions (\ref{term1four}),   (\ref{term2four}) or  (\ref{SgaugeAphi1}), if these terms  become not regularized.
For example, let us consider   the expression  (\ref{term1four}).

If we perform  the shift of variables $p \rightarrow p+k$ and $q \rightarrow q-k$  in the term  proportional to $  \frac{1}{\Lambda^2} A^{\mu}_{\alpha} (p)  A^{\nu}_{\beta} (q)  $,  its  UV behavior becomes worse.
This is because,  the  divergent function $\rho_1^{-2} (k)$, is now  only multiplied by the product of the fields $ A^{\mu}_{\alpha} (p+k)  A^{\nu}_{\beta} (q-k)$, and not by any  smeared field  $\phi$  function of the variable $k$ which adds the multiplicative factor  $\rho_1^2 (k)$.
Having in mind the rules outlined before, one can convince oneself that  the expressions (\ref{term1four}), (\ref{term2four}) and (\ref{SgaugeAphi1}) are indeed regularized if the propagators of the gluons and of the  $\varphi$ fields (\ref{smearphi1}) are rapidly decreasing functions in momentum space.
Notice that the contributions  to the action (\ref{Sinv2}) of the terms in  the second brace of (\ref{term2four}) can be regularized without any constraints on the  gluons and $\varphi$  propagators, knowing that the product of cutoff functions $\rho_2 (\rho_1^{-1} -\rho_2 ) $ is  a rapidly decreasing function of $k^2$ in momentum Euclidean space.

The contribution to  the action (\ref{Sinv2}) of its pure YM part  (\ref{puregauge1}) and of the  first,  third and last terms  of   (\ref{Sinv4}), are respectively given in  momentum space by the following expressions,
\beqa
\label{puregauge2}
S_{YM}&=& -\frac{1}{2g^2}  \int \dk ~\rho_1^{-2} (k) (k^2g_{\mu \nu}  -k_{\mu} k_{\nu})A^{\mu}_{ \alpha} (k) A^{\nu}_{ \alpha} (-k)  \nonumber \\
&& +\frac{i}{g^2}C_{ \alpha \beta\gamma} \int\dk \ddp  ~\rho_1^{-1} (k)\rho_2 (k)k^{\nu} A^{\mu}_{\alpha} (k) A_{\mu \beta} (p) A_{\nu \gamma} (-k-p) \nonumber \\
&&-   \frac{1}{4g^2} C_{ \alpha \beta\alpha' } C_{ \gamma \delta\alpha' }  \int\dk \ddp \dq ~\rho_2^2 (k) A^{\mu}_{\alpha} (p) A^{\nu}_{\beta} (k-p) A_{\mu\gamma} (q) A_{\nu \delta} (-k-q)
\eeqa
\beqa
\label{term2action}
S_{GG}&=&  \frac{1}{g^2} \Tr  \int \dk \ddp \dq ~\rho_2^2 (k) \bigg  \{ - (pqg_{\mu \nu}-p_{\nu}q_{\mu}) A^{\mu}_{\alpha} (p)  A^{\nu}_{\beta} (q) \Big [ \frac{1}{\Lambda^2} (\phi^+(p-k)T^{\alpha}  \nonumber \\
&&  +T^{\alpha} \phi(k-p))   (  \phi^+(k+q )T^{\beta} +T^{\beta}\phi(-k-q))   + \frac{2}{\Lambda^3}   \int  \ddp' ( \phi^+(p-k)T^{\alpha}  \nonumber \\
&& +  T^{\alpha}\phi(k-p))  \phi^+(p') T^{\beta} \phi(p'-k-q)   +\frac{1}{\Lambda^4} \int \ddp' \dq' \phi^+(p')T^{\alpha} \phi(p'+k-p) \nonumber \\
&& \times \phi^+(q')T^{\beta}  \phi(q'-k-q)\Big ] +\int  \dr p_{\nu}A^{\mu}_{\alpha} (p) A^{\nu}_{\beta} (q) A_{\mu \gamma}(r)\Big [  \frac{2i}{\Lambda^2} (\phi^+(p-k)T^{\alpha}  \nonumber \\
 && +T^{\alpha} \phi(k-p)) ( \phi^+(k+q+r) [T^{\beta},T^{\gamma}] +[T^{\beta},T^{\gamma}] \phi(-k-q-r))  \nonumber \\
&&   +  \frac{2i}{\Lambda^3}  \int  \ddp' (\phi^+(p-k)T^{\alpha}   +T^{\alpha} \phi(k-p))  \phi^+(p')[T^{\beta},T^{\gamma}]  \phi(p'-k-q-r)  \nonumber \\
&& +  \frac{2i}{\Lambda^3}  \int  \ddp' \phi^+(p')T^{\alpha} \phi(p'-p-k) (\phi^+(q-k+r)[T^{\beta},T^{\gamma}]  +[T^{\beta},T^{\gamma}] \nonumber \\
&& \times \phi(k-q-r))  +  \frac{2i}{\Lambda^4}  \int  \ddp' \dq'  \phi^+(p') T^{\alpha} \phi(k+p'-p)\phi^+(q')  \nonumber \\
&&  \times [T^{\beta},T^{\gamma}] \phi(q'-k-q-r) \Big ]  +  \int \dr \ds A^{\mu}_{\alpha} (p) A^{\nu}_{\beta} (q) A_{\nu \gamma}(r) A_{\mu \delta}(s)  \nonumber \\
&&  \times \Big [  \frac{1}{\Lambda^2}( \phi^+(p+q-k)T^{\alpha}T^{\beta} + T^{\alpha}T^{\beta}  \phi(k-p-q) )(\phi^+(k+r+s) [T^{\delta},T^{\gamma}]  \nonumber \\
&& + [T^{\delta},T^{\gamma}]  \phi(-k-r-s) )  +  \frac{2}{\Lambda^3} \int  \ddp' ( \phi^+(p+q-k)T^{\alpha}T^{\beta}  \nonumber \\
&&+  T^{\alpha}T^{\beta}   \phi(k-p-q))  \phi^+(p') [T^{\delta},T^{\gamma}] \phi(p'-k-r-s)  +\frac{1}{\Lambda^4} \int \ddp' \dq' \phi^+(p')  \nonumber \\
&&  \times T^{\alpha}  T^{\beta} \phi(k+p'-p-q)  \phi^+(q') [T^{\delta},T^{\gamma}]\phi^+(q'-k-r-s)  \Big ]  \bigg \}
\eeqa 
\beqa  
\label{term4action}
S_{GF}&=&  \frac{2}{g^2} \Tr  \int \dk \ddp ~\rho_1^{-1} (k) \rho_2 (k) \bigg  \{-(kpg_{\mu \nu}-k_{\mu}p_{\nu}) A^{\mu}_{\alpha} (p)  A^{\nu}_{\beta} (k) \Big [ \frac{1}{\Lambda} (\phi^+(k+p)T^{\alpha}  \nonumber \\ 
&& +T^{\alpha} \phi(-k-p)) T^{\beta} +  \frac{1}{\Lambda^2}  \int  \dq \phi^+(k+p+q)T^{\alpha}\phi(q)T^{\beta}  \Big ] +  \int  \dq k_{\mu} A^{\mu}_{\alpha} (p)  A^{\nu}_{\beta} (q) \nonumber \\
&& A_{\nu \gamma}(k) \Big [ \frac{i}{\Lambda} (\phi^+(p+q+k)[T^{\alpha},T^{\beta}] + [T^{\alpha},T^{\beta}]\phi(-k-p-q) )T^{\gamma}  \nonumber \\
&& +  \frac{i}{\Lambda^2}  \int  \ddp' \phi^+(p') [T^{\alpha},T^{\beta}] \phi(p'-k-p-q)  T^{\gamma} \Big ]  \bigg \}
 + \frac{2}{g^2} \Tr  \int \ddp \dq \dr  \rho_2^2 (q+r)  \nonumber \\
&& \bigg  \{ p_{\nu} A^{\mu}_{\alpha} (p)  A^{\nu}_{\beta} (q)A_{\mu \gamma}(r) \Big [ \frac{i}{\Lambda}(\phi^+(p+q+r)T^{\alpha}  +T^{\alpha} \phi(-p-q-r))  \nonumber \\
&& +  \frac{i}{\Lambda^2}  \int  \ddp' \phi^+(p')T^{\alpha} \phi(p'-p-q-r) \Big ][T^{\beta},T^{\gamma}]  +  \int  \ds  A^{\mu}_{\alpha} (p)  A^{\nu}_{\beta} (q)A_{\mu \gamma}(r)\nonumber \\
&&A^{\nu}_{\delta} (s) \Big [ \frac{1}{\Lambda} \phi^+(p+q+r+s) T^{\alpha}T^{\delta} + T^{\alpha}T^{\delta} \phi(-p-q-r-s) )\nonumber \\
&&-  \frac{1}{\Lambda^2}  \int  \ddp'   \phi^+(p') T^{\alpha} T^{\delta} \phi(p'-p-q-r-s)\Big][T^{\beta},T^{\gamma}]  \bigg \}
\eeqa 
\beqa  
\label{term6action}
S_{GH}&=&  \frac{2}{g^2} \Tr  \int \dk \ddp \dq ~ \rho_2 (k) (\rho_1^{-1} (k)- \rho_2 (k)) \bigg  \{A^{\mu}_{\alpha} (p) \Big [ \frac{i}{\Lambda^3} (kpq_{\mu} -qpk_{\mu}) (\phi^+(p-k)T^{\alpha} \nonumber \\
&&+ T^{\alpha} \phi(k-p))  \phi^+(q)  \phi(q-k) +\frac{i}{\Lambda^4}  \int  \ddp' (kpp'_{\mu} -pp'k_{\mu})  \phi^+(q) T^{\alpha}  \phi(k+q-p)\nonumber \\
&&\times\phi^+(p') \phi(p'-k) \Big ] + (kpg_{\mu \nu}-k_{\mu}p_{\nu}) A^{\mu}_{\alpha} (p)  A^{\nu}_{\beta} (q)  \Big [ \frac{1}{\Lambda^2} (\phi^+(p-k)T^{\alpha} + T^{\alpha} \phi(k-p))\nonumber \\
&&\times(\phi^+(k+q)T^{\beta} + T^{\beta}\phi(-k-q)) +  \frac{1}{\Lambda^3}  \int  \ddp' (\phi^+(p-k)T^{\alpha} + T^{\alpha} \phi(k-p))\nonumber \\
&& \times \phi^+(p') T^{\beta} \phi(p'-k-q)  +  \frac{1}{\Lambda^3}  \int  \ddp' \phi^+(p') T^{\alpha} \phi(k+p'-p)(\phi^+(k+q) T^{\beta}\nonumber \\
&&+ T^{\beta}   \phi(-k-q) )  +  \frac{1}{\Lambda^4}  \int  \ddp' \dq' \phi^+(p') T^{\alpha} \phi(k+p'-p)\phi^+(q') T^{\beta}  \phi(q'-k-q)  \Big ] \nonumber \\
&&+  A^{\mu}_{\alpha} (p)  A^{\nu}_{\beta} (q)  \int  \ddp' (k_{\mu}p'_{\nu} -k_{\nu}p'_{\mu}) \Big [ \frac{1}{\Lambda^3} (\phi^+(p+q-k) T^{\alpha}T^{\beta} +  T^{\alpha}T^{\beta}\phi(k-p-q))\nonumber \\
&&\times \phi^+(p') \phi(p'-k)+   \frac{1}{\Lambda^4}  \int   \dq' \phi^+(q') T^{\alpha} T^{\beta}\phi(k+q'-p-q)\phi^+(p')   \phi(p'-k) \Big ] \nonumber \\
&&  -i  \int  \dr k_{\mu} A^{\mu}_{\alpha} (p)  A^{\nu}_{\beta} (q) A_{\nu\gamma} (r)  \Big [ \frac{1}{\Lambda^2}(\phi^+(p+q-k)[T^{\alpha},T^{\beta}] + [T^{\alpha},T^{\beta}]\phi(k-p-q) ) \nonumber \\
&& \times (\phi^+(k+r)T^{\gamma} +T^{\gamma} \phi(-k-r) )  + \frac{1}{\Lambda^3}\int  \ddp'(\phi^+(p+q-k)[T^{\alpha},T^{\beta}] + [T^{\alpha},T^{\beta}]\nonumber \\
&&\times \phi(k-p-q) )\phi^+(p')T^{\gamma} \phi(p'-r-k)  + \frac{1}{\Lambda^3}\int  \ddp' \phi^+(p')[T^{\alpha},T^{\beta}]\phi(k-p+p'-q)\nonumber \\
&& \times(\phi^+(k+r)T^{\gamma} + T^{\gamma} \phi(-k-r) ) + \frac{1}{\Lambda^4}\int  \ddp' \dq'\phi^+(p')[T^{\alpha},T^{\beta}]\phi(k-p+p'-q)\nonumber \\
&& \times \phi^+(q')T^{\gamma}\phi(q'-k-r) \Big ]\bigg \}
\eeqa 
Since the cutoff function $ \rho_2$ and  the product of cutoff functions $\rho_2 \rho_1^{-1} $ are all    rapidly decreasing functions of $k^2$ in momentum Euclidean space, the contributions  to  the  action (\ref{Sinv2}) of the terms (\ref{term2action}),  (\ref{term4action}) and  (\ref{term6action}) are readily seen to be  all regularized.

In summary, all the contributions to the action described in this appendix are regularized, if  the cutoff function entering  the   smeared field  $\phi$  (\ref{smearphi1}) behaves at least as   (\ref{cutofphi1}), and if both the propagators of the gluons and the $\varphi$ fields are rapidly decreasing functions  in Euclidean momentum space in the IR and UV domains.
Moreover the shift of momenta of the   smeared field  $\phi$ are allowed in all  terms contributing to the action  (\ref{Sinv2}), apart  in some terms of the expressions (\ref{term1four}), (\ref{term2four}) and (\ref{SgaugeAphi1}).
\myappendix{Appendix D } 
\appD
After integration  over the boson fields $\varphi$, only the proper diagrams with internal  $\varphi$ bosons lines must be considered.
Regarding the structure of the action $S_{Reg}$ (\ref{fullaction1}) in momentum space, the vertices entering in such proper diagrams  $\Gamma$ can be classified in two sets.

Let $\Gamma^{1i}_{A\varphi}$ be the  set of vertices  given  by the parts   (\ref{term1four}), (\ref{term2four}), (\ref{SgaugeAphi1}),  (\ref{term2action}), (\ref{term4action}) and (\ref{term6action}) of the  action (\ref{fullaction1}).
In these vertices the derivative act in general  on the   boson fields $\varphi$.
If $N^{1i}_A$ is the number of gluon lines attached to the vertex   $\Gamma^{1i}_{A\varphi}$, the order of the derivative acting on  $\varphi$, is by inspection at most $4-N^{1i}_A$ .
Notice that  there are no derivatives acting on the $\varphi$  fields in a  set of vertices given by  the parts  (\ref{SgaugeAphi1}) and   (\ref{term2action})  of the  action (\ref{fullaction1}).
Thus if $N^{1i}_{\varphi}$ is the number of $\varphi$  fields attached   to the vertex  $\Gamma^{1i}_{A\varphi}$, by power counting the contribution of such a vertex 
to the UV behavior of the proper diagram  $\Gamma$ is
\beq
\label{vertex1}
\Gamma^{1i}_{A\varphi} \sim \Lambda^{4 -N^{1i}_A} ( \Lambda^{-1})^{N^{1i}_{\varphi}},
\eeq
where $\Lambda$ is the UV cutoff scale.
The part $S_{\pi}$ (\ref{massechi3}) of the action (\ref{fullaction1}) gives the second set  $\Gamma^{2i}_{\varphi}$ of vertices.
If  $N^{2i}_{\varphi}$ is the number of $\varphi$  fields attached   to the vertex  $\Gamma^{2i}_{\varphi}$, the contribution of this  vertex to the UV behavior of  $\Gamma$ is 
\beq
\label{vertex3}
\Gamma^{2i}_{\varphi} \sim  M^2 ( \Lambda^{-1})^{N^{2i}_{\varphi}-2}.
\eeq
Now suppose that the proper vertex  $\Gamma$ is composed of $L$ loops, with $m$ vertices of type $1$, $n$ vertices of type $2$ and $I_{\varphi}$ internal lines.
Since in the UV domain the propagators of the fields $\varphi$ behave like $M^{-2}$  (\ref{propsigma4}), in this domain  the proper vertex  $\Gamma$ scales like
\beq
\label{vertex4}
\Gamma \sim   \Lambda^{4L}  \Lambda^{[ 4m -\sum_{i \in S_{1m}}N^{1i}_A -(\sum_{i \in S_{1m}}N^{1i}_{\varphi} +  \sum_{i \in S_{2n}}N^{2i}_{\varphi}) +2n ]} M^{2n} M^{-2I_{\varphi}},
\eeq
where $S_{1m}$ and  $S_{2n}$ are respectively a subset of $m$ and $n$ indices referring to the vertices of type $1$ and $2$.
The total number $V$ of vertices and  the total number $I_{\varphi}$ of internal lines pertaining to the  proper vertex  $\Gamma$ (\ref{vertex4}) are respectively given by
\beq
\label{topology1}
V=m+n,
\eeq
and
\beq
\label{topology2}
I_{\varphi}=\frac{1}{2}\big(\sum_{i \in S_{1m}}N^{1i}_{\varphi} +  \sum_{i \in S_{2n}}N^{2i}_{\varphi} \big).
\eeq
Knowing that the number  $L$ of loops is related to  $V$ and  $I_{\varphi}$ \cite{ZUBER} by
\beq
\label{topology3}
L=I_{\varphi} -V +1,
\eeq 
the expression (\ref{vertex4}) can be reduced to
\beq
\label{vertex5}
\Gamma  \sim  \big  (\frac{ \Lambda^2}{M^2}\big  )^{I_{\varphi}-n} \Lambda^{-k}.
\eeq
Here the integer $k$ is given in terms of the  total  number $N_{A}=  \sum_{i \in S_{1l}}N^{1i}_A$ of external gluon lines pertaining to the proper vertex  $\Gamma$ (\ref{vertex4}) by
\beq
\label{topology4}
k= N_A -4,
\eeq
and the dimension of  $\Gamma$ (\ref{vertex5}) is exactly that of $ \Lambda^{-k}$.
The values of $k$ can be discussed according to the values of  $N_{A}$.

If $N_{A} >4$,  (\ref{topology4}) shows that $k$ is strictly positive.
Then,  by the dimensional argument, when all the external momenta scale with  $\Lambda$, the proper vertex $\Gamma$ (\ref{vertex5}) is suppressed by a power of $\Lambda$.
If $N_{A} \leq 4$,  only   the two, three and four effective gluons vertices must be considered.
At a fixed loop order, these effective vertices are a sum of proper vertices given by (\ref{vertex5}).
Therefore, from (\ref{vertex5}) and (\ref{topology4}), the  polarization operator,  whose dimension is that of a mass squared, seems to  behave at worst like $ \big  (\frac{ \Lambda^2}{M^2}\big  )^{I_{\varphi}-n} \Lambda^2$ in the UV domain.
However from the dimensional ground, Lorentz invariance and  gauge invariance of the theory,  this tensorial operator is transverse and then    behaves  in the UV domain like $ \big  (\frac{ \Lambda^2}{M^2}\big  )^{I_{\varphi}}  \mathcal {O} (\Lambda ^0)$.
This means that for the two gluons effective vertex, the minimal value of $k$ is not $-2$, but zero.
The behavior of the three  gluons effective vertex, is  at first sight $\big  (\frac{ \Lambda^2}{M^2}\big  )^{I_{\varphi}} \Lambda$.
Since the dimension of this vectorial operator is that of a mass, the Lorentz invariance imposes that this dimension is that of the external gluon  momenta.
In that case the minimal value of $k$ is in fact zero.
For the four  gluons effective vertex under consideration, the minimal value of $k$  (\ref{topology4}) is zero.

\end{document}